\documentclass[prd,nofootinbib,preprint,floatfix]{revtex4}

\usepackage{amsmath,amssymb}
\usepackage[normalem]{ulem}
\usepackage{graphicx}
\usepackage{array}
\usepackage{color}

\newcolumntype{C}{>{~$}c<{$~}}
\newcolumntype{R}{>{~$}r<{$~}}

\def\lsim{\raise0.3ex\hbox{$\;<$\kern-0.75em\raise-1.1ex
\hbox{$\sim\;$}}}
\def\gsim{\raise0.3ex\hbox{$\;>$\kern-0.75em\raise-1.1ex
\hbox{$\sim\;$}}}

\DeclareMathAlphabet{\mathsc}{OT1}{cmr}{m}{sc}

\definecolor{red}{cmyk}{0,1,1,0.4}

\begin{document}
\preprint{\vbox{%
\hbox{\bf YITP-SB-05-31}}}
\preprint{hep-ph/0511093}

\vspace*{.25in}
\title{Effects of Environment Dependence of Neutrino Mass versus
Solar and Reactor Neutrino Data}

\author{M.C.~Gonzalez-Garcia}
\email{concha@insti.physics.sunysb.edu}
\affiliation{
IFIC, Universitat de Val\`encia - C.S.I.C., Apt 22085, 
E-46071 Val\`encia, Spain \\
C.N.~Yang Institute for Theoretical Physics,
SUNY at Stony Brook, Stony Brook, NY 11794-3840, USA}
\author{P.C.~de Holanda} 
\email{holanda@fma.if.usp.br}
\author{R. Zukanovich Funchal}
\email{zukanov@if.usp.br}
\affiliation{
Instituto de F\'{\i}sica, Universidade de S\~ao Paulo, 
 C.\ P.\ 66.318, 05315-970 S\~ao Paulo, Brazil}

\begin{abstract}
In this work we study the phenomenological consequences of the
environment dependence  of neutrino mass on solar and reactor neutrino
phenomenology. We concentrate on mass varying neutrino scenarios in
which the enviroment dependence is induced by Yukawa interactions of
a light neutral scalar particle which couples to neutrinos and
matter. Under the assumption of one mass scale dominance, we perform a
global analysis of solar and KamLAND neutrino data which depends on 4
parameters: the two {\it standard} oscillation parameters, $\Delta
m^2_{0, 21}$ and $\tan^2\theta_{12}$, and two new coefficients which
parameterize the environment dependence of the neutrino mass.  We find
that, generically, the inclusion of the environment dependent terms
does not lead to a very statistically significant improvement on the
description of the data in the most favoured MSW LMA  (or LMA-I)
region.  It does, however, substantially improve the fit in the
high-$\Delta m^2$ LMA (or LMA-II) region which can be allowed at
98.9\% CL.  Conversely, the analysis allow us to place stringent
constraints on the size of the environment dependence terms which can
be translated on a bound on the product of the effective
neutrino-scalar ($\lambda^\nu$) and matter-scalar ($\lambda^N$) Yukawa
couplings, as a function of the scalar field mass ($m_S$) in these
models, $ |\lambda^\nu \, \lambda^N| \, \left(\frac{ 10^{-7}\,
\text{eV}}{m_S}\right)^2 \leq 3.0 \times 10^{-28}$ (at 90\% CL) .
\end{abstract}
\maketitle
\section{Introduction}
\label{sec:intro}
The possibility of environment dependence (ED) of the effective neutrino
mass was first proposed as a possible solution to the solar neutrino
deficit by Wolfenstein~\cite{wolfenstein}. In the Standard 
Model, the vector part of the standard charged current  and 
neutral current  neutrino-matter interactions contribute to 
the neutrino evolution equation as an energy independent 
{\sl potential} term. The potential is proportional to the 
electron density and it has different sign for neutrinos and
anti-neutrinos, even in the CP conserving case. 
It gives rise to the well-known Mikheyev-Smirnov-Wolfenstein (MSW)
effect~\cite{wolfenstein,ms} which is crucial in the interpretation of
the solar neutrino data.

In most neutrino mass models, new sources of ED of
the effective neutrino mass arise as a natural feature due to the presence
of non-standard neutrino interactions with matter~\cite{review}.  If
the new interaction can be cast as a neutral or charged vector
current, it will also contribute as an energy independent potential to
the neutrino evolution equation. The phenomenological implications of
such non-standard interactions in neutrino oscillations have been
widely considered in the literature~\cite{nsi,nsi2,nsi3}.

New physics in the form of Yukawa interactions of neutrinos and matter
with a neutral light scalar particle modify the {\sl kinetic} part of the
neutrino evolution equation.  Such neutral scalar interactions induce
a dependence of the neutrino mass on the environment~\cite{Sawyer}
which is energy independent and has the same sign for neutrinos and
anti-neutrinos. Recently, this form of ED of the neutrino mass has
received renewed attention after Ref.~\cite{dark1} discussed the
possibility that such mass varying neutrinos (MaVaNs) can behave as a
negative pressure fluid which contributes to the origin of the cosmic
acceleration. This scenario establishes a connection between
neutrino mass and dark energy with interesting cosmological
consequences~\cite{dark1,Peccei,Pas}.

In the MaVaNs scheme presented in  Ref.~\cite{dark1}, the neutrino
mass arises from the interaction with a scalar field, the acceleron,
whose effective potential changes as a function of the neutrino
density. As a consequence, the neutrino mass depends on the neutrino
density in the medium. A subsequent work, Ref.~\cite{dark2},
investigated the possibility that neutrino masses depend on the
visible matter density as well. Such a dependence would be induced by
non-renormalizable operators which would couple the acceleron to the
visible matter.  This form of ED of the neutrino
mass could also lead to interesting phenomenological consequences for
neutrino oscillations~\cite{dark2,Zurek,Barger,mymavas}.

For solar neutrinos, in Ref.~\cite{mymavas} it was shown that,
generically, due to the dependence of the neutrino mass on the
neutrino density, these scenarios establish a connection between the
effective $\Delta m^2$ in the Sun and the absolute neutrino mass
scale. Due to this effect, the description of solar neutrino data
worsens for neutrinos with degenerate masses.  On the other hand, for
hierarchical neutrino masses the dominant effect is the dependence 
of the neutrino mass on the visible matter
density.  In Ref.~\cite{Barger} it was shown that for some particular
values of the scalar-matter couplings this effect can improve
the agreement with solar neutrino data.

In this article we investigate the characteristic effects of the
dependence of the neutrino mass on the matter density for solar
neutrinos and reactor antineutrinos.  We perform a combined analysis
of solar~\cite{chlorine,sagegno,gallex,sk,sno,sno05} and KamLAND
data~\cite{kamland} in these scenarios.  Our results show that: $(i)$
the inclusion of the ED terms can lead to certain
improvement of the quality of the fit in the most favour LMA-I region
 for well determined values of the new parameters, (in agreement with
the result of Ref.~\cite{Barger}), but this improvement does not hold
much statistical significance; $(ii)$ the inclusion of these effects,
can substantially improve the quality of the fit in the high-$\Delta
m^2$ (LMA-II) region which can be allowed at 98.9\% CL; $(iii)$
generically, the combined analysis of solar and KamLAND data results
into a constraint on the possible dependence of the neutrino mass on
the ordinary matter density.

In Sec.~\ref{sec:forma}, we introduce the general theoretical 
framework which we will consider in this paper. In Sec.~\ref{sec:solarosc},  
we discuss how the ED modifies neutrino oscillations 
in matter.  In Sec.~\ref{sec:const}, we examine how these modifications can 
affect the current allowed solar neutrino oscillation parameter region  
and establish the constraints that solar and reactor neutrino data can 
impose on the couplings to the scalar field.  
Finally, in Sec.~\ref{sec:disc}, we discuss our results and summarize 
our conclusions.
 
\section{Formalism}
\label{sec:forma}
For the sake of concreteness, we consider here an effective 
low energy model containing the Standard Model particles plus
a light scalar ($\phi$) of mass $m_S$  which couples very weakly  
both to neutrinos ($\nu_i$) and the matter fields $f=e,n,p$. 

The Lagrangian takes the form
\begin{equation}
{\cal L}=
\sum_i \bar\nu_i\left(i\slash\!\!\!\partial\,\,- m^0_i\right)\nu_i+
\sum_f \bar f\left(i\slash\!\!\!\partial\,\,- m^0_f\right)f +
\frac{1}{2} \left[\phi \left(\partial^2- m_S^2\right)\phi\right]
+\sum_{ij} \lambda^\nu_{ij} \bar\nu_i \nu_j \phi
+\sum_{f} \lambda^f \bar f f \phi\;, 
\label{eq:lag}
\end{equation}
where $m^0_i$ are the {\sl vacuum mass} that the neutrinos would have in the
presence of the cosmic neutrino background. $\lambda^{\nu}_{ij}$ and
$\lambda^f$ are, respectively, the effective neutrino-scalar and
matter-scalar couplings.  We have written a Lagrangian for Dirac
neutrinos but equivalently it could be written for Majorana neutrinos.

In a medium with some additional neutrino background (either
relativistic or non-relativistic) as well as non-relativistic matter
(electrons, protons and neutrons), neutrinos acquire masses which obey
the following set of integral equations
\begin{eqnarray}
m_{ij}(r)&=&m^0_i \delta_{ij} - M_{ij}(r), \nonumber \\
M_{ij}(r)&=&\frac{\lambda^\nu_{ij}}{m_S^2} \left(\sum_f \lambda^f n_f(r)
+\sum_a {\lambda^\nu}_{aa} 
\int \frac{d^3 k}{(2\pi)^3} 
\frac{M_{aa}}{\sqrt{k^2 + M_{aa}^2}} f_a(r,k)\, \right). 
\label{eq:numass}
\end{eqnarray}
$n_f(r)$ is the number density for the  fermion $f$,  
and  $f_a(r,k)$ is the sum of neutrino and antineutrino ``$a$'' 
occupation numbers for momentum $k$ in addition to the 
cosmic background neutrinos.

In the context of the dark energy-related MaVaNs models of
Ref~\cite{dark1,dark2} the scalar $\phi$ would be the acceleron --
with mass in the range $m_S\sim 10^{-6}$--$ 10^{-8}$ eV --
which, when acquiring a non-vanishing expectation value,  
$\langle\phi\rangle$, gives a contribution to the neutrino mass.
This in turn implies that the acceleron effective potential receives a
contribution which changes as a function of the neutrino density, so that 
\begin{equation}
\lambda^\nu = \left.
\frac{\partial m_\nu}{\partial\phi}\right|_{\langle\phi\rangle}\; .
\label{eq:lnu}
\end{equation}
$\lambda^f$, the effective low energy couplings of the acceleron to
visible matter, come from non-renormalizable operators which couple 
the acceleron to the visible matter, such as might arise from
quantum gravity. They can be parametrized as
\begin{equation} 
\lambda^f= \tilde \lambda^f  \left(\frac{m_f}{M_{Pl}}\right)  , 
\label{eq:lmat}
\end{equation}
where $M_{Pl}$ is the Plank scale. 
Tests of the gravitational inverse square law require 
$\lambda^n, \lambda^p \lesssim 10^{-21}$ ~\cite{Adelberger} for any
scalar with $m_S\gtrsim 10^{-11}$ eV. 

The results in Eq.(\ref{eq:numass}) and Eq.(\ref{eq:lnu}) correspond
to the first order term in the Taylor expansion around the present epoch
background value of $\phi$. In general, for the required flat
potentials in these models, one needs to go beyond first order and the
neutrino mass is not linearly proportional to the number density of
the particles in the background.  The exact dependence on the
background densities is function of the specific form assumed for the
scalar potential. This is mostly relevant for the neutrino density
contribution to the neutrino mass.  

It has been recently argued~\cite{zaldarriaga} that, generically,
these models contain a catastrophic instability which occurs when
neutrinos become non-relativistic. As a consequence the acceleron
coupled neutrinos must be extremely light. Thus, in what follows we
assume the vacuum neutrino masses to be hierarchical
\begin{equation} 
0=m^0_1<m^0_2<m^0_3\; .
\label{eq:hier}
\end{equation}

For solar neutrinos of hierarchical masses, as discussed in 
Ref.\cite{mymavas,Barger}, the dominant
contribution to the neutrino mass is due to the matter background density. 
Correspondingly, we neglect the contribution to the neutrino mass 
from the background neutrino density and we concentrate on the matter 
density dependence:
\begin{equation}
M_{ij}(r)=\frac{\lambda^\nu_{ij}}{m_S^2} \sum_f \lambda^f n_f(r) \; .
\label{eq:mrmat}
\end{equation}
This is very similar to the scenario considered in Ref.\cite{Barger}.

Finally, let's mention that by assuming Eq.(\ref{eq:lag}) we do not
consider the possibility of additional light mixed sterile neutrinos 
which may appear in some specific realizations 
of  MaVaNs scenarios~\cite{Fardon:2005wc} and which can lead to other 
interesting effects in oscillation neutrino phenomenology and 
cosmology~\cite{Fardon:2005wc,Barger:2005mh,Weiner:2005ac}.

\section{Effects in Solar Neutrino Oscillations}
\label{sec:solarosc}
The minimum joint description of atmospheric~\cite{atm}, K2K~\cite{k2k}, 
solar~\cite{chlorine,sagegno,gallex,sk,sno,sno05} 
and reactor~\cite{kamland,chooz} data requires that all the 
three known neutrinos take part in the oscillations.  
The mixing parameters are encoded in the $3\times 3$ lepton mixing 
matrix which can be conveniently parametrized in the standard form 
\begin{equation}
    U=\begin{pmatrix}
    1&0&0 \cr
    0& {c_{23}} & {s_{23}} \cr
    0& -{s_{23}}& {c_{23}}\cr\end{pmatrix}
    \begin{pmatrix}
    {c_{13}} & 0 & {s_{13}}e^{i {\delta}}\cr
    0&1&0\cr 
    -{ s_{13}}e^{-i {\delta}} & 0  & {c_{13}}\cr 
\end{pmatrix} 
\begin{pmatrix}
    c_{21} & {s_{12}}&0\cr
    -{s_{12}}& {c_{12}}&0\cr
    0&0&1\cr
\end{pmatrix}
\end{equation}
where $c_{ij} \equiv \cos\theta_{ij}$ and $s_{ij} \equiv
\sin\theta_{ij}$. 

According to the current data, the neutrino mass squared 
differences can be chosen so that 
\begin{equation} \label{eq:deltahier}
    \Delta m^2_\odot = \Delta m^2_{21} \ll 
|\Delta m_{31}^2|\simeq|\Delta m_{32}^2|
=\Delta m^2_{\rm atm}.
\end{equation}
As a consequence of the fact that $\Delta m^2_{21}/\vert \Delta
m^2_{31}\vert \approx 0.03$, for solar and KamLAND neutrinos, the
oscillations with the atmospheric oscillation length are completely
averaged and the interpretation of these data in the neutrino
oscillation framework depends mostly on $\Delta m^2_{21}$,
$\theta_{12}$ and $\theta_{13}$, while atmospheric and K2K neutrinos
oscillations are controlled by $\Delta m^2_{31}$, $\theta_{23}$ and
$\theta_{13}$.  Furthermore, the negative results from the CHOOZ
reactor experiment~\cite{chooz} imply that the mixing angle connecting
the solar and atmospheric oscillation channels, $\theta_{13}$, is
severely constrained ($\sin^2\theta_{13} \leq 0.041$ at 3$\sigma$
\cite{haga}).  Altogether, it is found that the 3-$\nu$ oscillations
effectively factorize into 2-$\nu$ oscillations of the two different
subsystems: solar and atmospheric.

With the inclusion of the ED terms 
(Eq.~(\ref{eq:mrmat})) it is not warranted that such factorization
holds. We will assume that this is still the case and study
their effect on solar and KamLAND oscillations under the hypothesis of
one mass-scale dominance. Under this assumption, we parametrize the evolution 
equation as \cite{Barger}:
\begin{equation} 
i \frac{d}{dr} 
\left(\begin{array}{c} \nu_e \\
\nu_\mu     \end{array}\right)=\left[
\frac{1}{2E_\nu}
     \mathbf{U}_{\theta_{12}}
\left(\begin{array}{cc}
         M_1^2(r) &  M_3^2(r) \\
        M_3^2(r) & (m^0_2-M_2(r))^2
    \end{array}\right)
     \mathbf{U}_{\theta_{12}}^\dagger +
\left(    \begin{array}{cc}
        V_{\rm CC}(r) & ~0 \\
                    0 & ~0 
    \end{array}\right)\right]
\left(\begin{array}{c} \nu_e \\
\nu_\mu     \end{array}\right),
\label{eq:evol}
\end{equation}
where we have assumed the neutrinos to follow the hierarchy given  
in Eq.(\ref{eq:hier}). Here $V_{\rm CC}(r)= \sqrt{2} G_F n_e(r)$ is 
the MSW potential proportional to the electron number density  
$n_e(r)$ in the medium. $\mathbf{U}_{\theta_{12}}$ is the $2\times 2$ 
mixing matrix in vacuum parameterized by the angle $\theta_{12}$,
 and  $M_i(r)$ are  the ED contributions to the 
neutrino masses. 

In general, for given matter density profiles, Eq.~(\ref{eq:evol}) has
to be solved numerically. As discussed in Ref.\cite{Barger} in most of the 
parameter space allowed by KamLAND and solar data, for all practical
purposes, the transition is adiabatic and the evolution equation can be
solved analytically to give the survival probability
\begin{equation}
P_{ee}=\frac{1}{2}+\frac{1}{2}
\cos 2\theta^m_{0,12}\cos 2\theta_{12}\, ,
\label{eq:pee}
\end{equation}
where $\theta^m_{0,12}$ is the effective neutrino mixing angle 
at the neutrino production point $r_0$ in the medium, explicitly given by
\begin{equation}
\cos 2\theta^m_{0,12}=
\frac{(\Delta \tilde{M}_{21}^2(r_0)\cos 2\tilde{\theta}_{0,12}
-2 E_\nu V_{\rm CC}(r_0))}{
\sqrt{(\Delta \tilde{M}_{21}^2(r_0) 
\cos 2\tilde{\theta}_{0,12}-2 E_\nu V_{\rm CC}(r_0))^2+
(\Delta \tilde{M}_{21}^2(r_0) \sin 2\tilde{\theta}_{0,12})^2}},
\label{eq:costm}
\end{equation}
with
\begin{eqnarray}
\Delta \tilde{M}_{21}^2(r_0)=2\sqrt{M_3^4(r_0) +
\left(\frac{\Delta M^2_{21}(r_0)}{2}\right)^2}
\label{eq:deltatilde}
\\
\cos 2\tilde{\theta}_{0,12}=
\frac{\displaystyle\frac{\Delta M_{21}^2(r_0)}{2}\cos 2\theta_{21}-
M_3^2(r_0)\sin 2\theta_{12}}
{\sqrt{M_3^4(r_0)+\left(\displaystyle\frac{\Delta M^2_{21}(r_0)}{2}\right)^2}}
\label{eq:costhetatilde}
\end{eqnarray}
and where
\begin{equation}
\Delta M^2_{21}(r_0)=(m_2^0-M_2(r_0))^2-M_1^2(r_0).
\label{eq:M21}
\end{equation}

As discussed in the previous section, in general,  
$M_i(r)$  can be an arbitrary function of the background matter
density. For the sake of concreteness we will assume a dependence 
in accordance with  the results obtained in the linear approximation 
given in Eq.(\ref{eq:mrmat}). 
Furthermore, using Eq.(\ref{eq:lmat}), $\lambda^e\ll\lambda^n=\lambda^p
\equiv \lambda^N$, so we will parametrize these terms as:
\begin{equation}
M_i(r) = \alpha_i \, \left[\frac{\rho(r)}{\rm (gr/cm^{3})} \right]~~,
\label{eq:mral}
\end{equation}
where $\rho$ is the matter density, and  from Eq.(\ref{eq:mrmat}) we
find the characteristic value of the $\alpha$ coefficients to be 
\begin{equation}
\alpha \sim 4.8 \times 10^{23}\,  
\lambda^\nu \,  \lambda^N \, 
\left(\frac{10^{-7} \text{eV}}{m_S}\right)^2\, {\rm eV} \;.\label{eq:alfa}
\end{equation}

One must notice, however, that, as long as the transition is
adiabatic, the survival probability only depends on the value of
$M_i(r)$ at the neutrino production point.  Therefore it only depends
on the exact functional form of $M_i(r)$ via the averaging over the
neutrino production point distributions.

The survival probability for anti-neutrinos, $P_{\bar{e}\bar{e}}$,
which is relevant for KamLAND,  takes the form
\begin{equation}
P_{\bar{e}\bar{e}}=1-
\sin^2 2\tilde\theta^m_{0,12}\sin^2
\left(\frac{\Delta m^2_{KL} L}{2E_\nu}\right)\, ,
\label{eq:peekam}
\end{equation} 
where $\cos 2\tilde\theta^m_{0,12}$ is defined as in
Eq.~(\ref{eq:costm}) and $\Delta m^2_{KL}$ is the denominator of this
equation but replacing $V_{\rm CC}$ by $- V_{\rm CC}$ and assuming a
constant matter density $\rho \sim 3$ gr/cm$^3$, typical of the
Earth's crust.

\begin{figure}[ht]
\includegraphics[width=5.in]{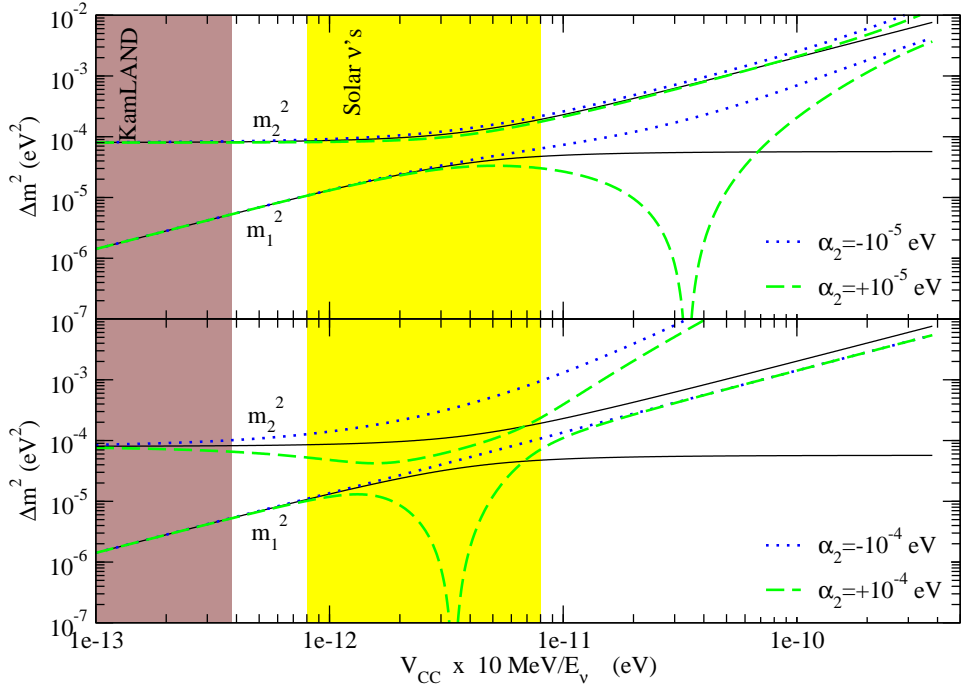}
\caption{Evolution of the neutrino mass eigenvalues in matter.
The solid lines represent the familiar MSW evolution. In this 
plot we have fixed $(m_2^0)^2=8 \times 10^{-5}$ eV$^2$, 
$\tan^2 \theta_{12}=0.4$ and $\alpha_1=\alpha_3=0$.
In the upper (lower) panel the dotted line represents 
the case $\alpha_2=-10^{-5}$ eV ($\alpha_2=-10^{-4}$ eV) and the 
dashed line the case $\alpha_2=+10^{-5}$ eV ($\alpha_2=+10^{-4}$) eV. 
The shaded regions correspond to typical values of $V_{\rm CC}$ in neutrino
production region in the center of the Sun for the Solar $\nu$'s region, 
and a constant Earth crust density
of 3 g/cm$^3$, with a proton density fraction of $Y=0.5$ and neutrino energies
varying from 3 to 10 MeV for the KamLAND region.}
\label{fig:mass}
\end{figure}

To illustrate the effects of the $\alpha$ coefficients, we show in
Fig.~\ref{fig:mass} the evolution of the mass eigenvalues $m_1$ and
$m_2$ in matter as a function of $V_{\rm CC} E_\nu$ for different values
of $\alpha_2$ (keeping $\alpha_1=\alpha_3=0$).  
As a reference, we also show in this figure the standard 
MSW evolution curve (solid line) for the oscillation parameters at 
 $\Delta m^2_{0,21}=(m_2^0)^2=8\times 10^{-5}$ eV$^{2}$ and  
$\tan^2\theta_{12}=0.4$, a point which  explains very well both solar
and KamLAND data.  From this plot we can appreciate that in the region
relevant to solar neutrino experiments
the evolution of the mass eigenvalues is not significantly different 
from the MSW one if $\vert \alpha_2 \vert \lsim$ 10$^{-5}$ eV. 
For larger values of
$\alpha_2$, such as $\vert \alpha_2 \vert =$ 10$^{-4}$ eV, we expect 
solar neutrinos to be affected.  On the other hand,
KamLAND data is very little affected by the  ED terms in this range 
of $\alpha_2$.

Figure 1 also illustrates a curious feature of these scenarios: 
the fact that it is possible  to find a value of the matter dependence 
term which exactly cancels  $\Delta m^2_{0,21}$.
It can be seen, directly from Eqs.~(\ref{eq:evol}), that
if for a particular point, $r_0$, in the medium, $m_2^0=M_2(r_0)$ and
$M_1=M_3=0$ ($\alpha_1=\alpha_3=0$) the lower mass eigenstate will be
zero while the higher one will be at the corresponding value of
$2V_{\rm CC}(r_0)E_\nu$.

Non-adiabatic effects in the Sun can also occur. In the region of 
relatively small $\alpha$ parameters, non-adiabaticity occurs when 
the parameters  are ``tunned''  to give a vanishing effective 
$\Delta m^2_{21}$ (the denominator of Eq.~(\ref{eq:costm})).  
This can be achieved, for example, 
with 
$\alpha_1=\alpha_2=0$ by solving the following set of equations inside the 
Sun:
\begin{eqnarray}
  (m^0_2)^2 \cos 2 \theta - 2 M^2_3(r) \sin 2 \theta & = &  2 E_\nu V_{\rm CC}(r), \\
  (m^0_2)^2 \sin 2 \theta + 2 M^2_3(r) \cos 2 \theta & = & 0.
\end{eqnarray}
It can be shown that for $\alpha_3= i\, 5.5 \times 10^{-5}$ eV, 
$\tan^2\theta=0.3$ and  $E_\nu=10$ MeV  this set of equations are fulfilled 
at $r/R_{\odot} \sim 0.027$, and the neutrinos would suffer a non-adiabatic 
transition on their way out of the Sun. However, in general for the small 
values of the $\alpha$ parameters discussed here, these non-adiabatic 
effects do not lead to a better description of the solar neutrino data. 

More generically, non-adiabatic effects occur for sufficiently large 
values of the $\alpha$ parameters so that one can disregard the standard 
MSW potential $V_{\rm CC}$ and the vacuum mass 
$m_2^0$ with respect to the matter density  mass dependent terms. 
In this case, as seen from Eq.~(\ref{eq:costm}), the mixing angle inside the 
Sun is constant and controlled by the $\alpha's$. At the border 
of the Sun, as the density goes to zero, the mixing angle is driven to its 
vacuum value in a strongly non-adiabatic transition. This scenario would 
be equivalent to a vacuum-like oscillation for solar neutrinos with
the ED of neutrino mass having to play a leading role 
in the interpretation of terrestrial neutrino experiments.  
We will leave the detailed analysis of the consequences of this type of 
non-adiabatic transitions for a future work. 

\section{Constraints from Solar and Reactor Neutrino Data}
\label{sec:const}
We present in this section the results of the global analysis of
solar and KamLAND 
for the specific realization discussed in the previous section. 
Furthermore, for simplicity, we will  restrict ourselves to the case 
$M_1(r)=m_1=0$. 

Details of our solar neutrino analysis have been described in previous
papers~\cite{oursolar,pedrosolar}. 
We use the solar fluxes from Bahcall and Serenelli (2005)~\cite{BS05}.
The solar neutrino data includes a total of 119 data points: 
the Gallium~\cite{sagegno,gallex} 
and Chlorine~\cite{chlorine} (1 data point)
radiochemical rates, the Super-Kamiokande~\cite{sk} zenith spectrum (44 bins),
and SNO data reported for phase 1 and phase 2.  The SNO
data used consists of the total day-night spectrum
measured in the pure D$_2$O (SNO-I) phase  (34 data points)~\cite{sno}, plus 
the full data set corresponding to the Salt Phase (SNO-II)~\cite{sno05}.
This last one includes the NC and ES event rates during the day and 
during the night (4 data points), and the CC day-night spectral data 
(34 data points). The analysis of the full data set of SNO-II is new
to this work. It is done by a $\chi^2$ analysis using the
experimental systematic and statistical uncertainties and their 
correlations presented in~\cite{sno05}, together with the theoretical 
uncertainties. In combining with the SNO-I data, only the theoretical 
uncertainties are assumed to be correlated between the two phases. 
The experimental systematics errors are considered to be uncorrelated 
between both phases.

For KamLAND, we directly adapt the $\chi^2$ map as given by the
KamLAND collaboration for their unbinned rate+shape
analysis~\cite{kamhomepage} which uses 258 observed neutrino candidate
events and gives, for the standard oscillation analysis, a
$\chi^2_{\rm min}$=701.35. The corresponding Baker-Cousins $\chi^2$
for the 13 energy bin analysis is $\chi^2_{\rm min}=13.1/11$ dof.  The
effect of MaVaN's parameters in KamLAND result was calculated assuming
a constant Earth density of 3 g/cm$^3$, and assuming that KamLAND are
sensitive to the vacuum value of $\Delta m^2_{0,21}$ and $\theta_{12}$
through an effective mass and mixing in a constant Earth density, 
respectivelly given by the denominator of Eq.~(\ref{eq:costm})
and Eq.~(\ref{eq:costhetatilde}), as described in Eq.~(\ref{eq:peekam}).

In presence of the ED contribution to the masses,
the analysis of solar and KamLAND data depends on four parameters: the two
{\it standard} oscillation parameters  
$\Delta m^2_{0, 21}=(m^0_2)^2$, and $\tan^2\theta_{12}$, 
and the two ED coefficients, 
$\alpha_2$, and $\alpha_3$. 
In this case, in order to cover the full CP conserving parameter 
space we allow the $\alpha$  parameters to vary in the range
\begin{equation}
 -\infty \leq \alpha_2\leq \infty 
\;\;\;\;\;\;\;\;\;\;\;
 -\infty \leq \alpha^2_3\leq \infty
\end{equation}

We find the best fit point  
\begin{eqnarray}
\tan^2\theta_{12}=0.49 ~ && ~\Delta m^2_{0, 21} = 
8.4\times10^{-5}~{\rm eV}^2 \nonumber \\  
\alpha_2=10^{-4}~ {\rm eV} ~ && 
~\alpha_3= i\, 2.0\times10^{-5}~ {\rm eV} ~. 
\label{eq:bestfit}
\end{eqnarray}

This is to be compared with the best fit point for no ED
of the neutrino mass, i.e., $\alpha_2=\alpha_3=0$ 
\begin{eqnarray}
\tan^2\theta_{12}=0.44 ~ && ~ \Delta m^2_{0,21} = 
7.9\times10^{-5}~{\rm eV}^2 
~.
\nonumber \\
\Delta\chi^2=2.5 &&
\label{eq:bestfitsm}
\end{eqnarray}
where $\Delta\chi^2$ is given with respect to the minimum 
at the best fit point in Eq.(\ref{eq:bestfit}).
Thus we find that although the inclusion of the ED
terms can lead to a small improvement of the quality of the fit 
(in agreement with the result of Ref.~\cite{Barger}),  
this improvement is not statistically very significant leading only 
to a decrease of  $\Delta \chi^2=2.5$ even at the cost of introducing 
two new parameters.

\begin{figure}[t]
\includegraphics[width=5.in]{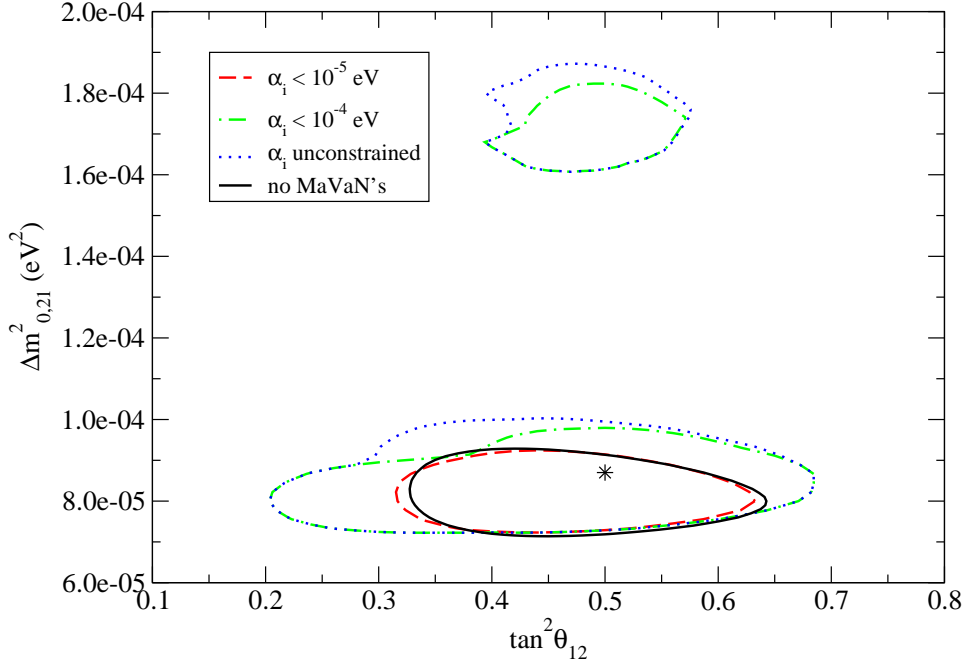}
\caption{Allowed regions from the global analysis of  
solar plus KamLAND data in the $(\Delta m^2_{0,21},\tan^2\theta_{12})$ 
parameter space at 3$\sigma$ CL (2dof). The best fit point at
$\tan^2\theta_{12}=0.5$ and $\Delta m^2_{0, 21} =
8.7\times10^{-5}~{\text{eV}}^2$ is represented by the star.  
The standard MSW allowed region is also shown for reference.}
\label{fig:oscregions}
\end{figure}

We show in Fig.~\ref{fig:oscregions} 
the result of the global analysis of solar data plus KamLAND data in the 
form of the allowed two-dimensional regions at 3$\sigma$ CL in the 
$(\Delta m^2_{21,0}, \tan^2\theta_{12})$ plane after marginalization 
over $\alpha_2$ and $\alpha_3$. The standard MSW allowed region is also 
showed for reference.
As seen in the figure, allowing for ED
of the neutrino masses enlarges only slightly the allowed range 
of $\Delta m^2_{21,0}$
and $\tan^2\theta_{12}$ in the LMA-I region. 
In contrast to the standard MSW analysis, where the limits on the 
mixing angle come basically from solar neutrinos, here it is KamLAND  
data that control the lower limits for the mixing angle.

Most interestingly, we  also find that the description of the solar 
data in the
high-$\Delta m^2$ (LMA-II) region can be significantly 
improved so there is a new allowed solution
at the 98.9\% CL. The best fit point in this region is obtained for
\begin{eqnarray}
\tan^2\theta_{12}=0.5 ~ && ~\Delta m^2_{0, 21} = 
1.75\times10^{-4}~{\rm eV}^2 \nonumber \\  
\alpha_2=1.3\times 10^{-4}~ {\rm eV} ~ && 
~\alpha_3= i\, 2.0\times10^{-5}~ {\rm eV}\; \nonumber \\
\Delta\chi^2=8.9 &&
\label{eq:bestfithigh}
\end{eqnarray}
While this region is excluded at more than 4$\sigma$ for standard MSW
oscillations, it is allowed at 98.9\% CL (2.55$\sigma$) in the presence of
environmental effects with $|\alpha_3|\leq 3.2 \times 10^{-5}$ and
$2.8\times 10^{-5}\leq \alpha_2 \leq 2.0 \times 10^{-4}$.  Basically
the CL at which this region is presently allowed is determined by
KamLAND data \cite{kamland} because the fit to the solar data cannot
discriminate between the LMA-I and LMA-II regions once the ED 
terms are included.  Clearly this implies that this solution will
be further tested by a more precise determination of the antineutrino
spectrum in KamLAND.

\begin{figure}[t]
\includegraphics[width=5.5in]{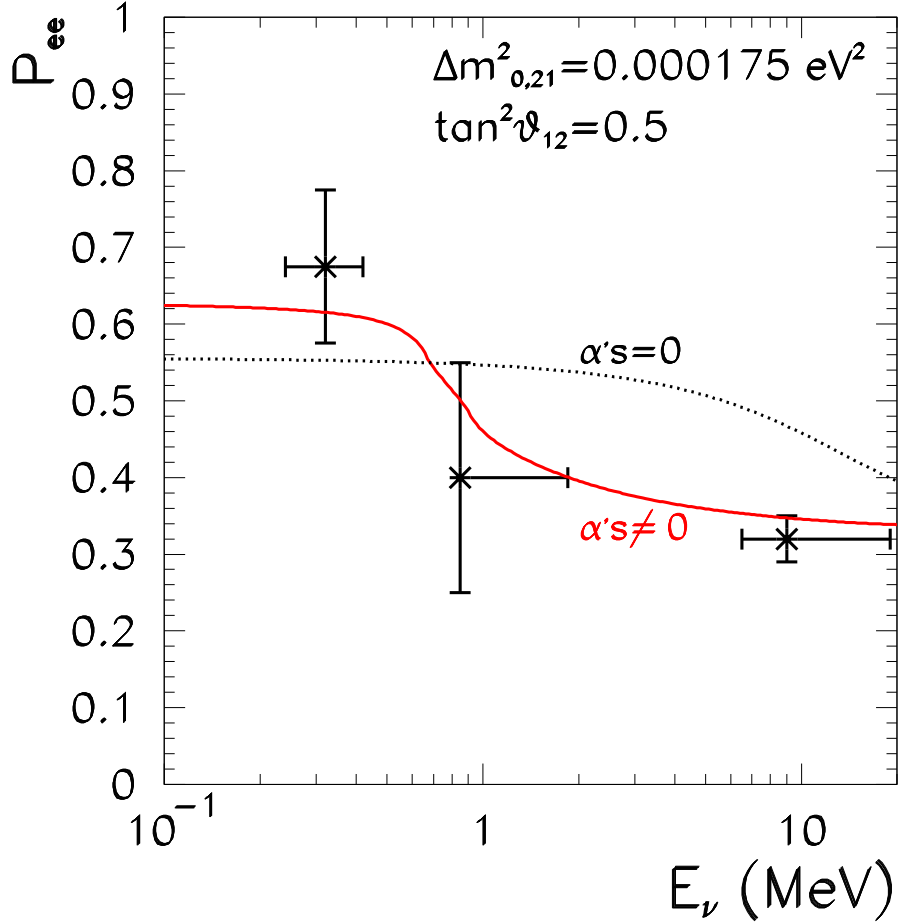}
\caption{$\nu_e$ survival probability in the Sun versus neutrino energy
for the best fit point in the high-$\Delta m^2$ region in the presence of
ED effects (Eq.~(\ref{eq:bestfithigh})).
The dotted line is the survival probability for conventional oscillations
($\alpha_i=0$) with the same values of $\Delta m^2_{21,0}$ and
$\theta_{12}$. These survival probabilities have been
obtained for neutrinos produced around $x=0.05$ as it is characteristic
of $^8$B and $^7$Be neutrinos.
The data points are the extracted average survival probabilities for the
low energy ($pp$) , intermediate energy ($^7$Be, $pep$ and CNO)
and high energy solar neutrinos ($^8$B and $hep$)
from Ref. \cite{Barger}.}
\label{fig:probhigh}
\end{figure}

We show in Fig.\ref{fig:probhigh} the survival probability for this
best fit point in the high-$\Delta m^2$ (LMA-II) region in the
presence of ED effects together with the
extracted average survival probabilities for the low energy ($pp$) ,
intermediate energy ($^7$Be, $pep$ and CNO) and high energy solar
neutrinos ($^8$B and $hep$) from Ref. \cite{Barger}. For comparison we
also show the survival probability for conventional oscillations
($\alpha_i=0$) with the same values of $\Delta m^2_{21,0}$ and
$\theta_{12}$.  From the figure it is clear that the inclusion of the
ED parameters, leads to an improvement on the
description of the solar data for all the energies being this more
significant for intermediate- and high-energy neutrinos.

On the contrary, unlike for the case of non-standard neutrino
interactions discussed in Ref.\cite{nsi2,nsi3}, the low-$\Delta m^2$
(LMA-0) region is still disfavoured at more than 3 sigma by the global
KamLAND and solar data analysis even in the presence of the new
``kinetic-like'' ED effects discussed here.  This
is due to the different energy dependence of the new physics effects
in the two cases. For the ``kinetic-like'' effects it is not possible to
suppress matter effects in the Earth for the high energy neutrinos (to
fit the SK and SNO negative results on the day-night asymmetry within
LMA-0) without spoiling the agreement of the survival probability at
intermediate energies with the result of the radiochemical
experiments.

\begin{figure}[t]
\includegraphics[width=5.5in]{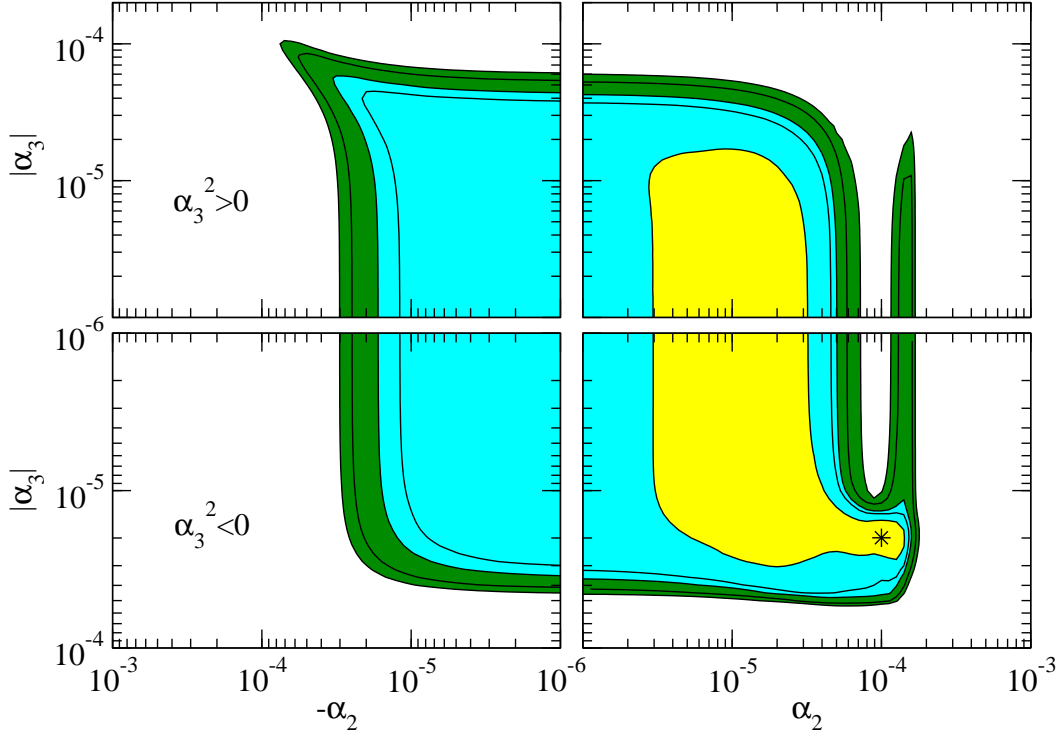}
\caption{Allowed regions from the global analysis of solar 
and solar plus KamLAND data in the $(\alpha_2,\alpha_3)$ parameter space.
The curves  correspond to 1$\sigma$, 90\%, 95\%, 99\% and 3$\sigma$ CL (2dof). 
The best fit point at $\alpha_2=10^{-4}~ {\text{eV}}$ and 
$\alpha_3= i\, 2.0\times10^{-5}~ {\text{eV}}$, represented by a star, 
is also shown.}
\label{fig:alfaregions}
\end{figure}

Conversely, the global analysis of solar and KamLAND data results into
the constraint of the possible size of the ED
contribution to the neutrino mass.  This is illustrated in
Fig.\ref{fig:alfaregions} where we show the result of the global
analysis in the form of the allowed two-dimensional regions in the
$(\alpha_2, \alpha_3)$ parameter space after marginalization over
$\Delta m^2_{0,21}, \tan^2\theta_{12}$.  The full regions correspond
to 1$\sigma$, 95\% and 3 $\sigma$ CL while the curves correspond to 90
and 99\% CL.  As seen in the figure, for CL $>$ 1.1$\sigma$ the
regions are connected to the $\alpha_2=\alpha_3=0$ case and they are
always bounded.  In other words, the analysis show no evidence of any
ED contribution to the neutrino mass and there is
an upper bound on the absolute values of the corresponding
coefficients.

Our previous discussion at the end of Sec.~\ref{sec:solarosc} on the
behavior of the mass eigenvalues shown in Fig.~\ref{fig:mass} when
$M_2 \rightarrow m_2^0$, is the reason behind the exclusion of the
narrow gulf around $\alpha_2 \approx 10^{-4}$ eV. In this region,
unless $\alpha_3$ is negative and not too small, it is not possible to
explain solar neutrino data.

\begin{figure}[ht]
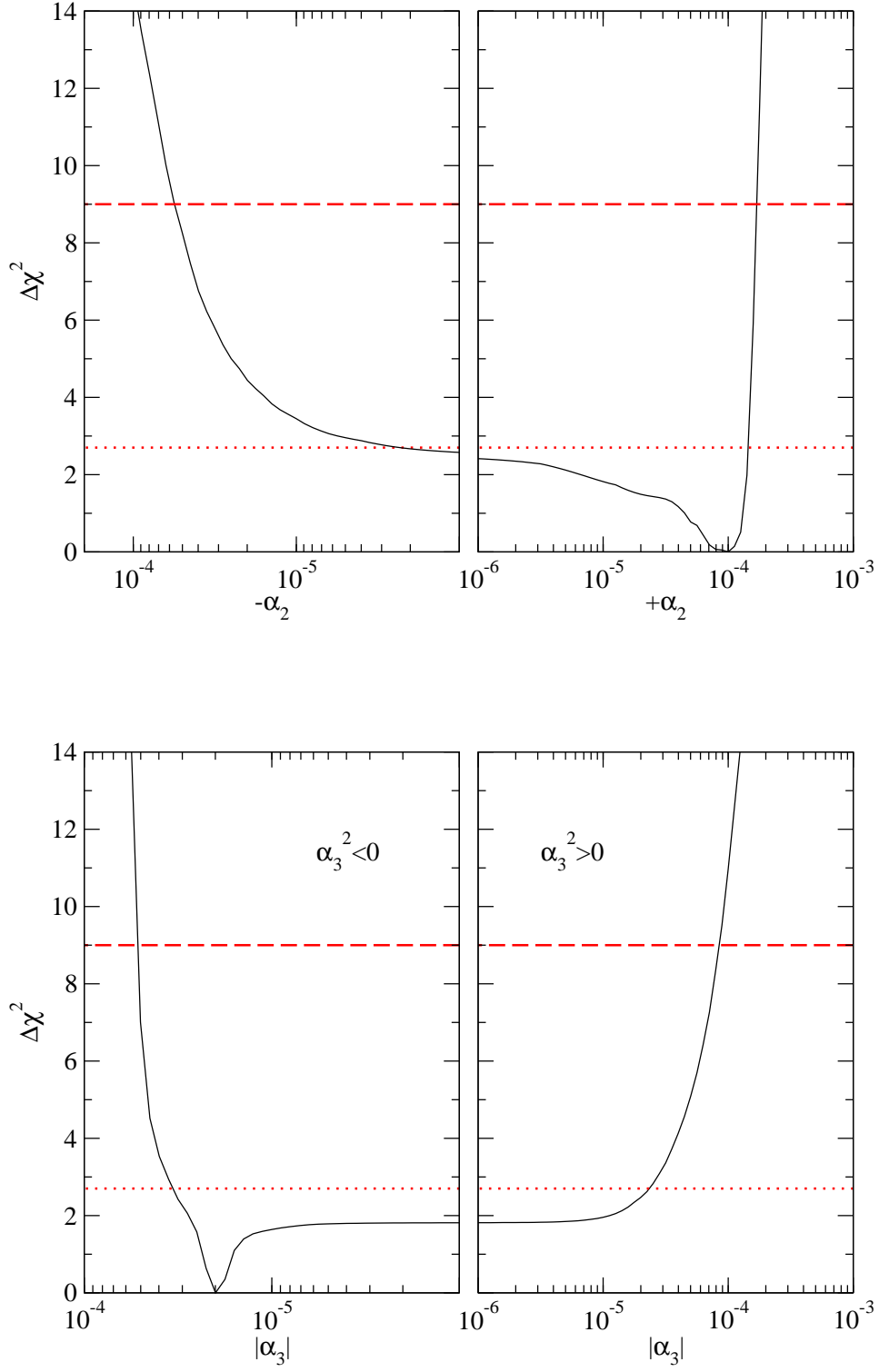

\includegraphics[width=5.in]{chi2minl1_fig5a.eps}
\vskip 1.7cm
\includegraphics[width=5.in]{chi2minl2_fig5b.eps}
\caption{Dependence of $\chi^2$ on the ED 
parameters $\alpha_2$ (upper panel) and $\alpha_3$ (lower panel) 
after marginalization over the other three parameters.}
\label{fig:chi2alfas}
\end{figure}

In order to quantify the bound on MaVaN's parameters, 
we display in Fig.\ref{fig:chi2alfas}
the dependence of the global $\chi^2$ on $\alpha_2$ ($\alpha_3$)
after marginalization over 
$\Delta m^2_{0,21}$, $\tan^2\theta_{12}$ and $\alpha_3$ ($\alpha_2$).
From the figure we read the following 90\% CL (3 $\sigma$),  
bounds (with 1dof)
\begin{eqnarray}
 -2.2\times 10^{-5} 
\leq  \alpha_2/{\rm eV}  \leq 1.4 \times 10^{-4} 
&&  ( -5.6\times 10^{-5} 
\leq  \alpha_2/{\rm eV}  \leq 1.7 \times 10^{-4} ) \\
|\alpha_3|/{\rm eV}
\leq 2.3  \times 10^{-5} &&
(|\alpha_3|/{\rm eV}
\leq 8.4  \times 10^{-5}) \;\; {\rm for}\;\; \alpha_3^2>0 \\
|\alpha_3|/{\rm eV}
\leq 3.4  \times 10^{-5} &&
(|\alpha_3|/{\rm eV}
\leq 5.2  \times 10^{-5}) \;\; {\rm for}\;\; \alpha_3^2<0 
\label{eq:limits}
 \end{eqnarray}

These bounds can be converted into a limit on the product of the 
characteristic effective neutrino-scalar and matter-scalar couplings. 
For example, at 90\% CL,
\begin{equation}
|\lambda^\nu \, \lambda^N | \, 
\left(\frac{10^{-7}\, \text{eV}}{m_S}\right)^2   \leq 3.0 \times 10^{-28} \, .
\label{eq:copl} 
 \end{equation}
We can compare this bound with those derived from tests of the 
gravitational inverse square law (ISL) which require the 
coupling of the scalar to nucleons 
$|\lambda^N| \lesssim 10^{-21}$ ~\cite{Adelberger} for any
scalar with $m_S\gtrsim 10^{-11}$ eV. Thus we find that if the scalar 
also couples to neutrinos with coupling
\begin{equation}
\lambda^\nu \gtrsim 3.0 \times  10^{-7} \left(\frac{m_S}{10^{-7}\, \text{eV}}\right)^2  
\end{equation}
the analysis of solar and KamLAND data yields a more restrictive 
constraint on the  matter-scalar couplings than ISL tests.

Finally, we want to comment on the possible model-dependence of these
results.  There are two main sources of arbitrariness in our
derivations: the choice of $m_1=M_1(r)=0$ and the assumption that the
$M_{i}$ are linearly dependent on the matter density. Indeed their
effect is the same: departing of any of these assumptions results into 
a different functional dependence of the effective neutrino masses with 
the point along the neutrino trajectory, $m_i(r)$.

As discussed in the previous section, the basic assumption behind 
our results is that neutrino evolution in matter is adiabatic.
As long as this is the case, the survival probability only depends on 
the value of the effective neutrino masses at the neutrino production 
point and the final results depend very mildly  on the exact functional 
form of $m_i(r)$. As a consequence the generic results will still be
valid: there will be a slight improvement on the quality of the fit in the
LMA-I region, there will be a substantial improvement of the quality
of the fit on the LMA-II region and generically the combined
analysis of solar and KamLAND data will result into a bound on the
strength of the new contributions. Of course, the exact numerical values 
of the corresponding ED couplings will be different. 
But the order of magnitude of the bound on the product of the Yukawa
couplings in Eq.~(\ref{eq:copl}) will hold.

\section{Discussion}
\label{sec:disc}

We have investigated the phenomenological consequences of a scalar
induced ED of the effective neutrino mass in the
interpretation of solar and reactor neutrino data.  For the sake of
concreteness, we consider an effective low energy model containing the
Standard Model particles plus a light neutral scalar ($\phi$) of mass $m_S$
which couples very weakly both to neutrinos ($\nu_i$) and the matter
fields $f=e,n,p$. This is described in Sec.~\ref{sec:forma} and its
consequences to neutrino oscillations in the Sun is discussed in
Sec.~\ref{sec:solarosc}.

Assuming the neutrino masses to follow the hierarchy 
$0=m^0_1<m^0_2<m^0_3$, we have performed a combined analysis of the  
 solar neutrino data (118 data points) and KamLAND (17 data points) in 
the context of this effective model. 
 Our analysis, which is described in Sec.~\ref{sec:const}, depends on 
4 parameters: the two {\it standard} oscillation parameters  
$\Delta m^2_{0, 21}=(m^0_2)^2$, and $\tan^2\theta_{12}$, 
and the two ED coefficients, $\alpha_2$, and $\alpha_3$.
We found  the best fit point at:
$\tan^2\theta_{12}=0.49$, 
$\Delta m^2_{0, 21} = 8.4\times10^{-5}~{\rm eV}^2$,
$\alpha_2=10^{-4}~ {\rm eV}$ and $\alpha_3= i\, 2.0\times10^{-5}~ {\rm eV}$.  
This point corresponds to a decrease of $\Delta\chi^2_{\text{min}}=-2.5$ 
in comparison to
the minimum in the case where no ED is considered. 
We conclude that in spite of the inclusion the two 
extra parameters, the improvement of the quality of the fit in 
the most favoured LMA-I MSW region is not very statistically significant.  

Most interestingly, we  find that the description of the solar data in the
high-$\Delta m^2$  (LMA-II) region can be significantly improved and 
there is a  new allowed solution at the 98.9\% CL. 
The best fit point in this region is obtained for
$\tan^2\theta_{12}=0.5$, 
$\Delta m^2_{0, 21} = 1.75\times10^{-4}~{\rm eV}^2$,
$\alpha_2=1.3 \times 10^{-4}~ {\rm eV}$ and 
$\alpha_3= i\, 2.0\times10^{-5}~ {\rm eV}$. 
This solution will be further tested by a more precise determination of 
the antineutrino spectrum in KamLAND.

In any case, our data analysis permit us to considerably limit the size of
the $\alpha$ coefficients (see Eq.~(\ref{eq:limits})) and from that
to derive a limit on the product of the effective neutrino-scalar and
matter-scalar Yukawa couplings depending on the mass of the 
scalar field (Eq.~(\ref{eq:copl})). In particular,  
for neutrino-scalar couplings $\lambda^\nu \gtrsim 3.0 
\times  10^{-7} (m_S/10^{-7}\text{eV})^2$  
our analysis of solar and KamLAND data yields a more restrictive 
constraint on the  matter-scalar couplings than gravitational ISL tests.

These scenarios will be further tested by the precise determination 
of the energy dependence of the survival probability  of solar 
neutrinos, in particular for low energies~\cite{lowe}.

\vspace{-0.3cm}
\begin{acknowledgments} 
\vspace{-0.3cm}
We thank C. Pe\~na-Garay for careful reading of the manuscript 
and comments. 
This work was supported by Funda\c{c}\~ao de Amparo \`a Pesquisa do
Estado de S\~ao Paulo (FAPESP) and Conselho Nacional de Ci\^encia e
Tecnologia (CNPq). MCG-G is supported by National Science Foundation
grant PHY-0354776 and by Spanish Grants 
FPA-2004-00996 and GRUPOS03/013-GV. R.Z.F. is also grateful to the 
Abdus Salam International Center for Theoretical Physics where the final 
part of this work was performed.

\end{acknowledgments}


\end{document}